\documentclass[11pt]{article}

\usepackage[utf8]{inputenc}
\usepackage{cite}  
\usepackage{graphicx}
\usepackage[margin=1.25in]{geometry}
\usepackage[usenames,dvipsnames]{color}
\usepackage{url}
\usepackage[colorlinks = true,
            linkcolor = blue,
            urlcolor  = blue,
            citecolor = blue,
            anchorcolor = blue]{hyperref}
            
\usepackage{orcidlink}
\usepackage{lineno}
\usepackage{wrapfig}

\def\Title#1{\begin{center} {\LARGE #1 } \end{center}}

\newenvironment{Abstract}{\begin{quotation} \begin{center}
                       ABSTRACT
     \end{center}\bigskip  }{\end{quotation}}

\newcommand\snowmass{\begin{center}\rule[-0.2in]{\hsize}{0.01in}\\\rule{\hsize}{0.01in}\\
\vskip 0.1in Submitted to the  Proceedings of the US Community Study\\ 
on the Future of Particle Physics (Snowmass 2021)\\ 
\rule{\hsize}{0.01in}\\\rule[+0.2in]{\hsize}{0.01in} \end{center}}

\def\beq{\begin{equation}}
\def\eeq#1{\label{#1}\end{equation}}
\def\eeqn{\end{equation}}

\newenvironment{Eqnarray}%
   {\arraycolsep 0.14em\begin{eqnarray}}{\end{eqnarray}}
\def\beqa{\begin{Eqnarray}}
\def\eeqa#1{\label{#1}\end{Eqnarray}}
\def\eeqan{\end{Eqnarray}}

\let\bar=\overbar

\def\lsim{\mathrel{\raise.3ex\hbox{$<$\kern-.75em\lower1ex\hbox{$\sim$}}}}
\def\gsim{\mathrel{\raise.3ex\hbox{$>$\kern-.75em\lower1ex\hbox{$\sim$}}}}

\def\del{\partial}
\def\Dslash{\not{\hbox{\kern-4pt $D$}}}
\def\dslash{\not{\hbox{\kern-2pt $\del$}}}
\def\pslash{\not{\hbox{\kern-2pt $p$}}}
\def\ETmiss{\not{\hbox{\kern-4pt $E$}}_T}

\def\Dlr{\mathrel{\raise1.5ex\hbox{$\leftrightarrow$\kern-1em\lower1.5ex\hbox{$D$}}}}

\def\MSB{{\bar{M \kern -2pt S}}}
\def\msb{{\bar{\scriptsize M \kern -1pt S}}}

\def\drb{{\bar{\scriptsize D \kern -1pt R}}}

\begin{document}


\snowmass{}

\Title{\textbf{Data Science and Machine Learning in Education}\\
\vspace{10pt}%
COMPF3 (Machine Learning)}



\begin{center} 

Gabriele~Benelli\,\orcidlink{}\textsuperscript{1},
Thomas~Y.~Chen\,\orcidlink{0000-0002-0294-3614}\textsuperscript{2},
Javier~Duarte\,\orcidlink{0000-0002-5076-7096}\textsuperscript{3},
Matthew~Feickert\,\orcidlink{0000-0003-4124-7862}\textsuperscript{4},
Matthew~Graham\,\orcidlink{}\textsuperscript{5},
Lindsey~Gray\,\orcidlink{}\textsuperscript{6},
Dan~Hackett\,\orcidlink{}\textsuperscript{7},
Phil~Harris\,\orcidlink{}\textsuperscript{7},
Shih-Chieh~Hsu\,\orcidlink{}\textsuperscript{8},
Gregor~Kasieczka\,\orcidlink{}\textsuperscript{9},
Elham~E.~Khoda\,\orcidlink{0000-0001-8720-6615}\textsuperscript{8},
Matthias~Komm\,\orcidlink{}\textsuperscript{10},
Mia~Liu\,\orcidlink{}\textsuperscript{11},
Mark~S.~Neubauer\,\orcidlink{0000-0001-8434-9274}\textsuperscript{4},
Scarlet~Norberg\,\orcidlink{}\textsuperscript{12},
Alexx~Perloff\,\orcidlink{0000-0001-5230-0396}\textsuperscript{13},
Marcel~Rieger\,\orcidlink{}\textsuperscript{10},
Claire~Savard\,\orcidlink{}\textsuperscript{13},
Kazuhiro~Terao\,\orcidlink{}\textsuperscript{14},
Savannah~Thais\,\orcidlink{}\textsuperscript{15},
Avik~Roy\,\orcidlink{0000-0002-0116-1012}\textsuperscript{4},
Jean-Roch~Vlimant\,\orcidlink{}\textsuperscript{5}
Grigorios~Chachamis\,\orcidlink{0000-0003-0347-0879} \textsuperscript{16}
\end{center}

\vspace{-18pt}
\begin{center}
\textbf{1}~Brown University, Providence, RI 02912, USA
\textbf{2}~Columbia University, New York, NY 10027, USA
\textbf{3}~University of California San Diego, La Jolla, CA 92093, USA
\textbf{4}~University of Illinois at Urbana-Champaign, Urbana IL 61801, USA
\textbf{5}~California Institute of Technology, Pasadena, California 91125, USA
\textbf{6}~Fermi National Accelerator Laboratory, Batavia, IL 60510, USA
\textbf{7}~Massachusetts Institute of Technology, Cambridge, MA 02139, USA 
\textbf{8}~University of Washington, Seattle, WA 98195, USA
\textbf{9}~Universit{\"a}t Hamburg, Institut f{\"u}r Experimentalphysik, 22761 Hamburg, Germany
\textbf{10}~European Organization for Nuclear Research, Geneva, Switzerland
\textbf{11}~Purdue University, West Lafayette, IN 47907, USA
\textbf{12}~University of Puerto Rico Mayag{\"u}ez, Mayag{\"u}ez, Puerto Rico
\textbf{13}~University of Colorado Boulder, Boulder, CO 80309, USA
\textbf{14}~SLAC National Accelerator Laboratory, Menlo Park, CA 94025, USA
\textbf{15}~Princeton University, Princeton, NJ 08544, USA
\textbf{16}~Laborat{\' o}rio de Instrumenta\c{c}{\~ a}o e F{\' \i}sica Experimental de Part{\' \i}culas (LIP), Lisboa, Portugal. 
\end{center}

\begin{Abstract}
\vspace{-4pt}
\noindent
\small
The growing role of data science (DS) and machine learning (ML) in high-energy physics (HEP) is well established and pertinent given the complex detectors, large data, sets and sophisticated analyses at the heart of HEP research. 
Moreover, exploiting symmetries inherent in physics data have inspired physics-informed ML as a vibrant sub-field of computer science research. 
HEP researchers benefit greatly from materials widely available materials for use in education, training and workforce development. 
They are also contributing to these materials and providing software to DS/ML-related fields. 
Increasingly, physics departments are offering courses at the intersection of DS, ML and physics, often using curricula developed by HEP researchers and involving open software and data used in HEP. 
In this white paper, we explore synergies between HEP research and DS/ML education, discuss opportunities and challenges at this intersection, and propose community activities that will be mutually beneficial. 
\end{Abstract}


\pagebreak

\section{Introduction}
\label{sec:introduction}

The particle physics research community has a strong background and involvement in educational activities. Not only do many of its practitioners come from universities and centers for education, but the community also provides training and educational resources to facilitate our science and convey its importance to members of the public and policy makers.

Particle physics holds a prominent role within academic curriculum at institutions of learning. 
There are compelling reasons for this prominent role, such as the fundamental nature of our science, fascinating historical development of our field, theoretical research that applies (and often develops) advanced mathematics, powerful applications such as cancer treatment, and high-visibility spin-off technologies such as the World Wide Web.

Data science and machine learning have an increasingly prominent role in our science, as is evident from any recent particle physics conference and this Snowmass process. In recent years, machine learning techniques for detector and accelerator control~\cite{StJohn:2020bpk}, data simulation~\cite{Butter:2020tvl}, parton distribution functions~\cite{Forte:2020yip}, reconstruction~\cite{Kasieczka:2019dbj,Duarte:2020ngm,Elkarghli:2020owr}, anomaly detection~\cite{Nachman:2020ccu,Alanazi:2021grv} and data analysis are increasing being applied to particle physics research. Recent reviews of these techniques applied to particle physics research can be found in~\cite{Albertsson:2018maf,Guest:2018yhq,PRESSCUT-H-2018-405,Bourilkov:2019yoi,Psihas:2020pby,Larkoski:2017jix,Carleo:2019ptp,Schwartz:2021ftp,Karagiorgi:2021ngt,Deiana:2021niw}. 
A ``living review'' aiming to provide a comprehensive list of citations for those in the particle physics community developing and applying machine learning approaches to experimental, phenomenological, or theoretical analyses can be found in~\cite{Feickert:2021ajf}.

\begin{wrapfigure}[16]{R}{0.5\textwidth}
  \vspace{-20pt}
  \begin{center}
    \includegraphics[width=0.48\textwidth]{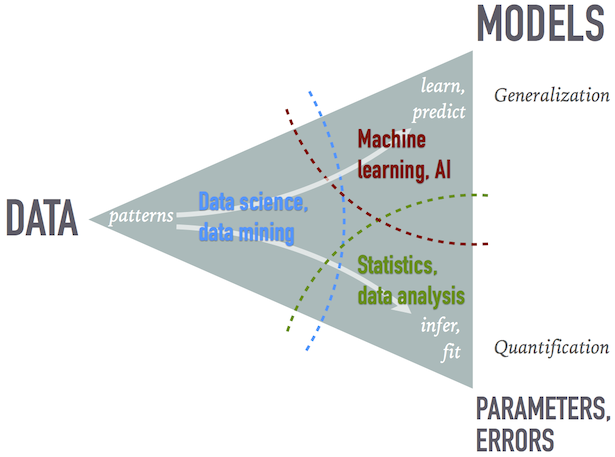}
  \end{center}
  \vspace{-20 pt}
  \caption{Simplified illustration of data science and machine learning in the context of data, models and statistical inference.}
\label{fig:DSML}
\end{wrapfigure}

There is no consensus on the precise definitions of data science and machine learning. 
For our purposes, we consider \textit{data science} to refer to scientific approaches, processes, algorithms and systems used to extract meaning and insights from data~\cite{Dhar} and \textit{machine learning} to refer to techniques used by data scientists that allow computers to learn from data. Machine learning is a subset of the field of artificial intelligence which aims to develop systems that can make decisions typically requiring human-level expertise, possessing the qualities of intentionality, intelligence and adaptability~\cite{shukla2013applicability}. 
Figure~\ref{fig:DSML} illustrates these relationships. 

Particle physicists increasingly collaborate with computer scientists and industry partners to develop ``physics-driven" or ``physics-inspired" machine learning architectures and methods.
However, the particle physics community in the U.S. has been generally slow to adopt data science and machine learning as formal components in educational curriculum. This situation is rapidly improving. The potential synergies between education and particle physics research in the areas of data science and machine learning motivates this study. 

In this Snowmass white paper, we explore some of the challenges and opportunities for data science and machine learning in education and suggest future directions that could benefit the particle physics community.

\section{Educational Pathways}
\label{sec:pathways}
There are many ways that particle physics researchers at all levels provide educational opportunities to students and other trainees. 
Mentoring in research activities is a key educational delivery method that provides an opportunity for professional development of early career individuals and bi-directional learning inherent in scientific research.  

In addition to traditional advising of undergraduate and graduate in thesis research, there are dedicated programs such as the NSF Research Experience for Undergraduates~\cite{reu}, masterclasses~(E.g. \cite{master}), and capstone project-based courses~(E.g. \cite{ucsdcapstone}). These programs are multidisciplinary and often involve direct involvement with industry which create strong opportunities for students to pursue careers within or beyond academia.

Particle physicists also develop curriculum at the intersection of data science, machine learning and physics for use in undergraduate and graduate-level courses in their home department(s). These curriculum might represent whole courses or specific teaching modules which could be used in multiple courses or other forms of educational delivery. A few examples are provided in Sec.~\ref{sec:examples}. 
These materials are often drawn from and feed into educational and training materials from particle physics research, such as schools (e.g.~\cite{CMSschool,LPCschool,CERNschool}), training events (e.g.~\cite{HSF_training,CODASHEP}), workshops (e.g~\cite{Katz:2020bvz,ML4Jets2020,physicsmeetsml,mlps2020,mlps2021}) and bootcamps (e.g.~\cite{USATLAS_Computing_Bootcamp_2019,USATLAS_Computing_Bootcamp_2020}).

\section{Examples of Curriculum from HEP researchers}
\label{sec:examples}

Particle physicists have been active in generating course curriculum and in many cases entire courses around the intersection of physics, data science and machine learning. 
Sec.~\ref{sec:HEP4EDU} and Sec.~\ref{sec:EDU4HEP} describe some of the motivating factors that compel particle physicists to devote time to educational aspects of data science and machine learning in physics.

In this section, we provide a few specific examples of courses developed by members of our community. This is a small sampling of courses and development efforts in data science and machine learning for education, included by the authors as a point of reference. There are many such courses developed by physicists outside of HEP for their respective physics departments (e.g.~\cite{Mehta-course,Coughlin-course}) from which we could learn a great deal in the spirit of interdisciplinary collaboration. 
In Sec.~\ref{sec:future}, we propose a more coordinated effort of discovery of such courses and the sharing of tools and experiences for those that have taught and/or developed such courses. Recently, several textbooks have also been developed specifically for teaching machine learning to (particle) physicists~\cite{DLforPR,AIforHEP,MLforPhysicists}.

\vspace{2mm}
\noindent\textit{\underline{Particle Physics and Machine Learning}} (UCSD)~\cite{duarte-course}: This is a course on particle physics developed by Javier Duarte and Frank W\"{u}rthwein as part of the UCSD Data Science Capstone sequence~\cite{ucsdcapstone}. 
The course centers around applying modern machine learning techniques to particle physics data.
The intended audience consists of fourth-year data science majors, who select among a set of topics for their capstone projects.
The course is split into two quarters.
The bulk of the first quarter focuses on the task of identifying Higgs boson decaying to bottom quarks. Specifically, the students are tasked with reproducing results in a recent ML and particle physics publication~\cite{Moreno:2019neq} on graph neural networks for the identification of boosted Higgs bosons decaying to bottom quarks.
They are also provided with weekly lectures on data science topics (by the lecturer)  as well as particle physics topics (by the domain mentors).
During the second quarter, students propose and execute a new project that extends the work of the previous quarter. Possible projects include studying the performance of different message passing and graph neural network structures, studying mass decorrelation strategies, applying explainable AI techniques (like layerwise relevance propagation) to the Higgs tagging task, comparing multiclassification to binary classification, and developing a network for Higgs boson jet mass regression.

\vspace{2mm}
\noindent\textit{\underline{Physics and Data Science}} (MIT)~\cite{harris-course}: This is a course developed by Phil Harris and aims to present modern computational methods by providing realistic examples of how these computational methods apply to physics research. Topics include: Poisson statistics, error propagation, fitting, data analysis statistical measures, hypothesis testing, semi-parametric fitting, deep learning, Monte Carlo simulation techniques, Markov-chain Monte Carlo, and numerical differential equations. The class format is a mixture of lectures by course faculty (and guest speakers to highlight the relevance of this work towards department research), recitations run by undergraduate and graduate students, and completion of three projects with a final presentation based on an extension of any one of the three projects. The projects include data analysis in gravitational waves, cosmic microwave background, and LHC jet physics using open data.

\vspace{2mm}
\noindent\textit{\underline{Introduction to Machine Learning}} (Princeton)~\cite{thais-course}:
This is a course developed by Savannah Thais and is primarily intended for non-computer science students who want to understand the foundations of building and testing an ML pipeline, different model types, important considerations in data and model design, and the role ML plays in research and society. Topics covered in lectures and exercises include conceptual foundations of ML, artificial neural networks, convolutional and recurrent neural networks, unsupervised learning, generative models and topics in AI ethics such as data bias, algorithmic auditing, predictive policing, inequitable utilization of algorithms, proposed regulation.

\vspace{2mm}
\noindent\textit{\underline{Data Analysis and Machine Learning Applications for Physicists}} (Illinois)~\cite{illinois-mla}:
This is a course developed by Mark Neubauer which aims to teach the fundamentals of analyzing and interpreting scientific data and applying modern machine learning tools and techniques to problems commonly encounters in physics research. 
The class format is a combination of lectures, homework problems that elaborate on topics from the lectures that give students hand-on experience with data, and a final project that students can choose from in the areas of particle physics and astrophysics. Topics covered include handling, visualizing and finding structure in data, adapting linear methods to nonlinear problems, density estimation, Bayesian statistics, Markov-chain Monte Carlo, variational inference, probabilistic programming, Bayesian model selection, artificial neural networks and deep learning. 

\subsection{Tools and Techniques}
Not surprisingly, the course examples just described utilize different tools and employ varied approaches to course delivery given independent and tailored development at their respective universities. In spite of this, there is a significant overlap in the tools and techniques used in these courses, which are briefly described in this section.

\vspace{2mm}
\noindent\textit{\underline{Software}}
These courses make extensive use of open source software, especially Python as a core language. 
Python is a very popular language for education given it is a high-level language with automatic memory management, simplicity of its syntax and readability of its code as compared with most other languages. Python is a binary platform-independent language such that it can be run on virtually any hardware platform and operating system. This aspect is important in an educational setting where students use a variety computing systems. Python is an open source project that is free to use, with an extensive ecosystem of code libraries for applications in science, engineering, data science and machine learning. The example courses described make extensive use of libraries used in scientific computing, mathematics, and statistics such as \texttt{NumPy}~\cite{harris2020array}, \texttt{Pandas}~\cite{reback2020pandas,mckinney-proc-scipy-2010}, \texttt{matplotlib}~\cite{Hunter:2007} and \texttt{seaborn}~\cite{Waskom2021}. \texttt{SciPy}~\cite{2020SciPy-NMeth} provides algorithms for optimization, integration, interpolation, eigenvalue problems, algebraic equations, differential equations, statistics and many other classes of problems. 
In terms of machine learning, these courses use the general purpose \texttt{scikit-learn} library~\cite{scikit-learn}, as well as deep learning frameworks such as \texttt{PyTorch}~\cite{NEURIPS2019_9015} and \texttt{TensorFlow}~\cite{abadi2016tensorflow}.

Python as a language for data science and machine learning has broad community support. 
Therefore, a key benefit of using Python in the classroom in terms of professional and skills development is that it is a language used extensively in real-world applications of data science and machine learning and widely used in industry. 
Of course, this is only the present landscape and it is anyone's guess as how it will change on the 5--10 year timescale.

\vspace{2mm}
\noindent\textit{\underline{Data}}
The specific data used in the example courses described in this paper vary according to the exact lessons and projects being taught. However, the courses generally made use of open scientific data sources when appropriate. This is especially relevant for the project components of these courses. For example, open data resources at the UCI Machine Learning Repository~\cite{UCIML}, Galaxy Zoo challenge data~\cite{GalaxyZoo}, and CERN Open Data Portal~\cite{CERNOpenData} were used for HEP and astrophysics students projects in the course at Illinois.

\vspace{2mm}
\noindent\textit{\underline{Tools}}
The use of Jupyter notebooks was a common aspect of the example course described in this section. 
Jupyter notebooks provide an interactive front-end to a rich ecosystem of python packages that support machine learning and data science tasks. They provide a means for students and instructors to create and share documents that integrate code, \LaTeX-style equations, computational output, data visualizations, multimedia, and explanatory text formatted in markdown into a single document. When hosted by a cloud-based server resource such as JupyterLab, using these notebooks has huge benefits for teaching, including removing the need to install any software locally or require any specific machine to be used by students~\cite{jupyterEdu}.

\vspace{2mm}
\noindent\textit{\underline{Course Materials}}
The reference materials used in the courses were education and training materials that are widely available in the public domain and enhanced by a significant amount of supplementary resources linked on the course pages. In the Illinois course, all materials are managed through a dedicated Github Organization. The students and course staff are all members of this organization, with different access levels to material (repositories) according to their role. Students each create a private repository which is how they submit their homework and final projects for grading.

\vspace{2mm}
\noindent\textit{\underline{Infrastructure}}
As with data used in the example courses, the  infrastructure utilized for course delivery varied according to the specific needs of the courses and institutional arrangements. In general, the courses used open source software in the Python language to implement scientific codes, and commonly used machine learning frameworks and libraries within Jupyter notebooks. The Python code that the students developed could be executed in a number of ways within these courses. For example, a common course environment could be generated by package management software (e.g. Anaconda~\cite{Anaconda}) or a Linux container service (e.g. Docker~\cite{Docker}). Another approach was to use an execution environment such as Google Colab~\cite{Colab} which allows anybody to write and execute arbitrary python code through the browser, and is especially well suited to machine learning, data analysis and education. More technically, Colab is a hosted Jupyter notebook service that requires no setup to use, while providing access free of charge to computing resources including GPUs~\cite{Colab}. 
In the Illinois course example, a custom Docker container~\cite{illinois-mla-docker} maintained by the course staff is launched onto commercial cloud resources which used to serve notebooks for the students using JupyterLab and provide computational resources to execute the code. 

\vspace{2mm}
\noindent\textit{\underline{Delivery}}
The primary methods of content delivery and active student engagement varied by course but generally involved a mixture of lectures by course faculty that included physics and data science pedagogy demonstrated through in-class live examples in Jupyter notebooks, recitation/discussion style activities involving hands-on interactive exercises, and projects. 
In the MIT course, guest speakers were invited to highlight the relevance of the pedagogy with ongoing research in the department. 
In the UCSD course, the students were actively involved in proposing and executing a new project that extends the work of the previous quarter.

\section{Opportunities}
\label{sec:opportunities}

Physics departments are increasingly offering curriculum to their undergraduate and graduate students at the intersection of physics, data science and machine learning. 
Particle physicists are increasingly interested in developing new courses at this intersection. For those so inclined, these courses provide opportunities for particle physicists to (1) describe synergies between modern machine learning research and particle physics research, (2) make connections with colleagues from other departments, (3) make connections within their own department in other research domains, (4) recruit students interested research at the intersection of machine and particle physics, and (5) learn the tools and techniques from data science and machine learning that can be applied to particle physics research. 

There are opportunities to take advantage of programs in education from federal agencies and engage with key organizations, such as the American Physical Society's (APS) Topical Group on Data Science (GDS)~\cite{APSGDS}. 
The APS GDS is focused on promoting research at the growing interface between physics and data science, spanning big data, machine learning, and artificial intelligence, with relevance to HEP and other scientific domains such as astronomy and materials science. 
The Data Science Education Community of Practice (DSECOP)~\cite{DSECOP}, a program funded by the APS Innovation Fund and led by the APS Group on Data Science (GDS), seeks to support physics educators in integrating data science in their courses. 
DSECOP achieves this through
\begin{itemize}
    \item A Slack community of physics educators and industry professionals to discuss data science education in physics courses. Specifically, conversation will be around challenges, opportunities, and cutting edge skills necessary for a wide-range of jobs.
    \item Workshops~\cite{DSECOP-workshops} to promote shared understanding and solidify the community
    \item Supporting DSECOP Fellows, a group of early career physicists (graduate students and postdocs) who receive modest stipends to develop and test data science education materials~\cite{DSECOP-github}.
\end{itemize}

The DSECOP Fellow program is a good example of how researches at an early career stage can be strongly involved in curriculum development. 
In several of the example courses described in Sec.~\ref{sec:examples}, students and postdocs from the instructors research group were involved in curriculum development and course delivery as part of their professional development. 
The same is true for several of the training and bootcamp events described in Sec.~\ref{sec:pathways}.

Large NSF Institutes such as the Institute for Research and Innovation in Software for High Energy Physics (IRIS-HEP)~\cite{IRISHEP}, AI Institute for Artificial Intelligence and Fundamental Interactions (IAIFI)~\cite{IAIFI}, and the Accelerated Artificial Intelligence Algorithms for Data-Driven Discovery Institute (A3D3)~\cite{A3D3} have HEP as a research driver and substantial efforts in education and training. 
The development of course curriculum for data science and machine learning is synergistic with particle physics research efforts within these and other institutes. 

\subsection{What does HEP research have to offer for ML/DS Education?}
\label{sec:HEP4EDU}

HEP research has much to offer education in data science and machine learning. Research in HEP has long required advanced, cutting-edge computing techniques, and physicists have historically contributed to the development of these methods. 
Over the past decades, there have been great advances in the data processing power of machine learning algorithms and these methods have quickly been adopted by physicists to address the unique timing, memory, and latency constraints of HEP experiments.

Our science typically involves analysis of large datasets generated by complex instruments at the frontier of scientific research. 
It is enabled by application of machine learning methods and data science tools which helps demonstrate the power and importance of these in a scientific research setting as well better understand their limitations. 
In recent times, cutting edge research in ML methods have been tested for their effectiveness and scalability in problems that interest high energy physics. 
For instance, generative models have found application in calorimeter simulation~\cite{paganini2018calogan, paganini2018accelerating}, graph neural networks (GNNs) have been explored for particle flow reconstruction~\cite{pata2021mlpf}, and jet classification~\cite{moreno2020interaction}. 
These applications serve as compelling evidence of wide range applicability of ML models for large, complicated datasets. Incorporating these exercises in ML pedagogy can enable the students to learn about the analytical and practical aspects of implementation of complex models that include hyperparameter optimization, data and model parallelization, uncertainty quantification, and model interpretation.

In short, in particle physics we have some of the most compelling scientific applications of data science and machine learning that involve very large and complex datasets. 

Particle physics is also impacting machine learning research and therefore machine learning education. The constraints of HEP experiments and known symmetries of physical systems create a rich environment for the development of novel and physics-informed machine learning (see for example~\cite{cranmer2020lagrangian,DBLP:journals/corr/abs-2109-13901,Liu:2021azq,Maiti:2021fpy}). There are even entire conferences and workshops dedicated to this intersection including the Microsoft Physics $\cap$ ML lecture series~\cite{physicsmeetsml} and the ML and the Physical Sciences workshop at NeurIPS~\cite{mlps2020,mlps2021}.

Exploration of machine learning models within the domain of high energy physics goes beyond usual regression and classification problems. The pursuit of discovery in physics requires explicit understanding of the causal relationship between the inputs and outputs of an analysis model and classical ML techniques that help understand such relationships- polynomial regression models, decision trees, and random forests for instance, have found numerous applications in physics problems, including parameterized cross-section estimation for novel physics models~\cite{roy2020novel} and event classification for $H \rightarrow \gamma\gamma$ channel in search of Higgs boson at the LHC~\cite{chatrchyan2013observation}. However, deep neural networks are becoming increasingly popular and showing improved performance over simpler ML techniques (see for instance \cite{10.21468/SciPostPhys.7.6.076} for comparison of classical and deep neural techniques for identifying longitudinal polarization fraction in same-sign $WW$ production). These complex, highly nonlinear models often comprising $\mathcal{O}(100k)$ parameters are extremely difficult to explain and often regarded as black box surrogates. Recent literature has focused towards explainable AI (xAI) and a number of methods~\cite{ribeiro2016model, lundberg2017unified, bach2015pixel, ying2019gnnexplainer, shrikumar2017learning, schlichtkrull2020interpreting} have been explored to identify importance of features and intermediate hidden layers in the context of a wide range of deep neural network models. Application of these methods in HEP research~\cite{turvill2020survey, mokhtar2021explaining} is quickly becoming popular, as model interpretability remains a crucial aspect to determine the relevance of physics insights for ML models. With a long history of using interpretable models for physics research, HEP research allows validation of xAI techniques for mainstream ML research and pedagogy.

\subsection{What does ML/DS Education have to offer HEP research?}
\label{sec:EDU4HEP}

From the very beginning of high energy collider experiments, data analysis and statistical techniques have been integral for HEP research that have led to important discoveries, precision measurements, and setting limits on search for physics beyond the standard model. 
Frequentist and Bayesian statistical models are ubiquitous in our treatment of uncertainties associated with finite size of MC simulation, detector response, particle and jet reconstruction~\cite{lyons2013bayes}. HEP research has, therefore, significantly benefited from pedagogical introduction to data science and statistics. 
As a result, recent particle physics workshops have arranged for dedicated theoretical and hands-on sessions to introduce concepts of statistics that play important role in everyday HEP research~\cite{cern-fnal-summer-school, tasi-2018}. 
A number of tutorials~\cite{barlow2019practical, Cowan:2773595} and textbooks~\cite{behnke2013data, lista2017statistical} have been written on this subject.
With the overwhelming emergence of ML techniques over the past decade in order to solve important questions in HEP research, this trend can be extrapolated to include ML training in the standard curriculum of HEP curricula and workshops. 

Materials developed to familiarize students with ML methods can benefit the community in two different ways. First, they enable the researchers of the future to make use of state-of-the art tools in data analysis techniques. Second, instructors and professors who develop these materials can benefit from these ventures by learning about the most recent developments in different areas of ML and CS and then applying those methods in their own research. This also opens up opportunities for multi-disciplinary projects such as FAIR4HEP~\cite{FAIR4HEP} and A3D3~\cite{A3D3}, that can symbiotically benefit research across different fields of study such as HEP, multimessenger astronomy and computational neuroscience. Such endeavors help make HEP research more visible to multiple communities, developing a broader interest among students and researchers in learning about and solving problems in HEP.

Finally, as all instructors know, from teaching assistant to professor levels, teaching a course really helps one understand at a deeper level the material being taught. This is true for traditional physics curriculum as well as courses at the intersection of physics, data science and machine learning. Teaching in this broader space makes us better researchers in particle physics as practitioners of these tools and methods, and keeping up with current developments (to some limited degree).

\section{Challenges}
\label{sec:challenges}

There a numerous challenges confronting particle physicists incorporating data science and machine learning into the physics curriculum. 
We touch on a few of them in this section and include some ideas for mitigating the risks associated with these challenges. 

\subsection{Student (and Instructor) Preparation}
It may be surprising to some HEP researchers, but coding is new for many physics students at the undergraduate or even graduate level. Many are unfamiliar with languages such as Python and tools such as Git and Jupyter notebooks. Also, certain subjects like linear algebra are required to understand machine learning even at an introductory level. Lack of preparation can lead to frustration, stress and possibly failure in these type courses. Fortunately, foresight by the instructor on enforcing appropriate prerequisites and the providing of ample resources (with examples) for coding in Python such as \textit{A Whirlwind Tour of Python} by Jake VanderPlas (O’Reilly Media, Inc) can go a long way to providing an platform for student success.
On the point of prerequisites however, one must realize that there are also equity, diversity, and inclusion considerations, in that underrepresented minority students may have less exposure to the necessary coding/software background.

\subsection{Rapid development}
It is difficult enough for HEP researchers to keep up with current developments and literature in one's own field of research let alone two or three. Data science and machine learning are currently moving at a very rapid pace which means that some tools and approaches become obsolete on timescale of years or even between semesters. 
It is important to keep up on literature within the \textit{HEP domain} on applications of data science and machine learning to HEP research. 
It is also useful to read CS and ML papers to understand the some of the current developments and penetrate domain jargon (as people outside of HEP try to do with our published work), but the field is moving too fast to keep current with all developments, some of which do not have an obvious application to our science. 
This is not to say we should look out for potential HEP applications of ML developments, but at least for undergraduate curriculum it is best to focus primarily reasonably established applications. 

\subsection{Distinction and Adoption}
We want to make sure that courses developed in the physics department distinguish themselves from other courses and are valued by the students they are designed to teach. 
Two suggestions are as follows:
\begin{enumerate}
\item {\bf Not trying to do too much}. Our strengths in HEP lie in the analysis and interpretation of large scientific data sets and physics-inspired AI. Leave the foundational AI pedagogy to the CS courses.
\item {\bf Balancing physics and ML pedagogy}. 
Remember that its a physics course taught in the Physics Department. 
It's best to use as many physics examples and datasets to support your instruction as possible. I.e. Classification of jets and galaxies over cats and dogs.
\end{enumerate}

\subsection{Career Development at the HEP/DS/ML Intersection}

Despite the demonstrated criticality and vibrancy of HEP research, from a physics background the path to a sustainable research career at the intersection of physics and ML is unclear at best. For early career researches interested in this intersection, this research interest breadth needs to be affirmed and nurtured. 
It is good to help make connections between early career HEP researchers with this interest and those that have transitioned from academic to industry or other pursuits. 
\subsection{Tools and Infrastructure}

We have discussed in this paper several successful examples of tools and infrastructure utilized in courses at the intersection of physics, data science and machine learning. 
Some challenges around tools and infrastructure include (1) uniformity of software environment for students, (2) keeping the primary focus on physics by minimizing prerequisite on coding skills, as discussed previously, (3) providing sufficient GPU resources to train deep networks often of of most interest, (4) hosting/caching of datasets for use in course projects and (5) maintaining a working software environment with modern DS/ML tools between course transitions. 
On the point of maintenance, it is not uncommon for the API for some software package to change between semesters and break notebooks that worked before, requiring care for maintenance and evolution of course materials. 

\section{Future Directions and Recommendations}
\label{sec:future}

In this section, we describe some future directions based on the experiences just described that would help HEP researchers interested in physics education that includes data science and machine learning pedagogy.

\subsection{Tools and Infrastructure}

To address prerequisites, a basic Python-based programming course for physicists can go a long way toward improving the foundation for students.

The most important consideration in terms of tools and infrastructure is to have these element work well without detracting from the learning environment. As a simplistic example, if I want to use a lamp to illuminate a room, I just want the lamp to be functional and the electrical infrastructure to work without being distracted with all the details of how electrical current arrives at the outlet (of course, that is interesting in other contexts). The same is true for a course in physics and machine learning. 

Software and tools to launch student code within notebooks on CPUs, GPUs, and other resources to study aspects of physics should be open, portable, robust and easy to make physics education using DS/ML most effective.

Containers are a great technology to provide custom, course-specific software and data environments for use by student on infrastructure. These custom containers can be hosted externally and launched on cloud-based services to execute student notebooks. 
Students just write code in notebooks with a common software environment and the backend resources are provisioned for cell executions. 
Further customization of the run-time environment is of course possible with the appropriate instructions in the notebook (e.g. via \texttt{pip install}).

All of these functions are available with existing technologies. However, it is important for universities to provide support to educators, who are most often not experts in these technologies, in maintaining a working environment of tools and infrastructure. 
Increased sharing of experiences with the tools and infrastructure used for education in data science and machine learning among researchers in HEP and other fields is strongly encouraged. 

\subsection{Role of Open Data and AI Models Adhering to FAIR Principles}

The FAIR principles (findable, accessible, interoperable, and reusable) were originally proposed to inspire scientific data management for reproducibility and maximal reusability~\cite{wilkinson2016fair}. 
These principles can be extended as guidelines to explore management and preservation of digital objects like research software~\cite{lamprecht2020towards} and AI models~\cite{katz2021working}. 
If data and models are preserved in accordance with the FAIR principles, they can be reliably reused by researchers and educators to reproduce benchmark results for both research and pedagogical purposes. 
For instance, the detailed analysis of FAIR and AI-readiness of the CMS $H(b\overline{b})$ dataset in Ref.~\cite{chen2022fair} has explained how the FAIR readiness of this dataset has been useful in building ML exercises for the course~\cite{duarte-course} offered at UCSD. 
In fact, making data and models FAIR helps understand their context, content, and format, enabling transparent provenance and reproducibility~\cite{samuel2020machine}. 
Such practices help with interpretation of AI models by allowing comparison of benchmark results across different models~\cite{katz2021working} and application of \textit{post-hoc} xAI methods. 
FAIR can facilitate education in ML in numerous ways, like interpretability, uncertainty quantification, and easy access to data/models.

The use of real data sets  communicates to students that they are doing publication-quality work at the interface of machine learning and physics~\cite{viviana}.
Course-based undergraduate research experiences (CUREs) in DS/ML and physics can also help foster a sense of identity and belonging in the field for students~\cite{cure}.

\subsection{Curation and Coordination of Educational Materials}

We recommend careful curation of materials for data science and machine learning in physics education and the open sharing of these materials. 
We also encourage a forum to openly share the experiences of this type of teaching and general discovery of who has developed and taught such courses at their institutions. Answering questions like:
\begin{itemize}
    \item What courses have been developed and delivered by our community?
    \item What open data is available for possible use in ML/DS education?
    \item What training/education/bootcamp/hackathon materials already exist?
\end{itemize}
How can we better collect and expose the above information for use in our community? 
The information is available but diffuse and therefore some coordination would make the sharing of knowledge and experiences much more efficient. 
This type of coordinated effort could make the sharing and improvement of projects utilizing HEP data more efficient.

\section{Conclusions}
\label{sec:conclusions}
We believe it will become ever more crucial that both our young and experienced researchers have a working understanding of data science and machine learning tools. We would like to see a continuation of community efforts towards raising he level of ML proficiency among current researchers by providing in-depth and innovative schools and other training events. It would also be advantageous to see more movement towards the addition of DS and ML studies within the physics curriculum at our educational institutions. 
All of this will take cooperation from the entire HEP community. We have described in this white paper some of the experiences from HEP researchers in ML education, outlined opportunities and challenges, and recommended future directions to make this area more efficient and effective for the HEP community. 

\def\thefootnote{\fnsymbol{footnote}}
\setcounter{footnote}{0}


\bibliographystyle{JHEP}
\bibliography{main}

\end{document}